# RadarRGBD : A Multi-Sensor Fusion Dataset for Perception with RGB-D and mmWave Radar

Tieshuai Song, Jiandong Ye, Ao Guo, Guidong He, Bin Yang

*Abstract*—Multi-sensor fusion has significant potential in perception tasks for both indoor and outdoor environments. Especially under challenging conditions such as adverse weather and low-light environments, the combined use of millimeter-wave radar and RGB-D sensors has shown distinct advantages. However, existing multi-sensor datasets in the fields of autonomous driving and robotics often lack high-quality millimeter-wave radar data. To address this gap, we present a new multi-sensor dataset—RadarRGBD. This dataset includes RGB-D data, millimeter-wave radar point clouds, and raw radar matrices, covering various indoor and outdoor scenes, as well as low-light environments. Compared to existing datasets, RadarRGBD employs higher-resolution millimeter-wave radar and provides raw data, offering a new research foundation for the fusion of millimeter-wave radar and visual sensors. Furthermore, to tackle the noise and gaps in depth maps captured by Kinect V2 due to occlusions and mismatches, we fine-tune an open-source relative depth estimation framework, incorporating the absolute depth information from the dataset for depth supervision. We also introduce pseudo-relative depth scale information to further optimize the global depth scale estimation. Experimental results demonstrate that the proposed method effectively fills in missing regions in sensor data. Our dataset and related documentation will be publicly available at：https://github.com/song4399/RadarRGBD.

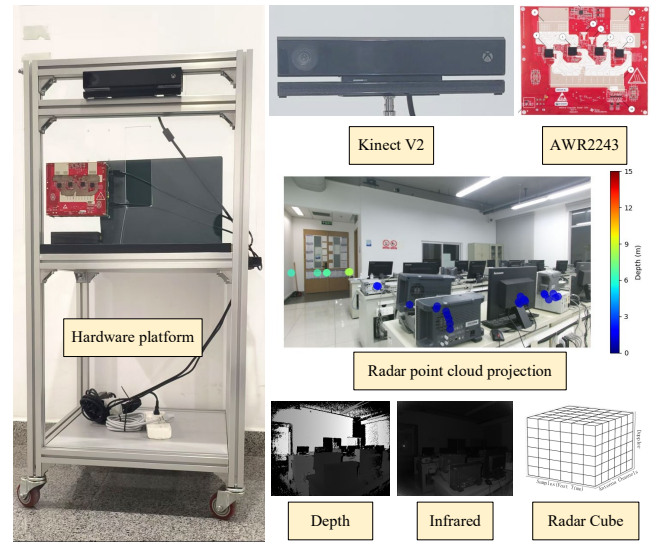

Figure 1. System composition

## I. INTRODUCTION

In recent years, multi-sensor fusion technology has been increasingly utilized in environmental perception systems, especially for tasks such as object detection, navigation, and localization in complex environments. The integration of millimeter-wave radar and RGB-D sensors has demonstrated distinct advantages in adverse weather conditions and low-light environments. A single sensor often fails to provide comprehensive information, and with the advancement of multi-sensor fusion algorithms, combining data from various sensors to achieve more accurate and robust perception capabilities has emerged as a critical research direction.

RGB-D sensors, such as Microsoft Kinect and Intel RealSense, are capable of real-time acquisition of 3D structural information of the environment. However, under challenging outdoor conditions, low-light environments, or when detecting distant objects, optical or infrared images are often susceptible to noise, lighting changes, and occlusions, leading to significant degradation in environmental perception accuracy. In contrast, millimeter-wave radar, by transmitting and receiving high-frequency electromagnetic waves, can accurately measure the distance and velocity of objects without being affected by lighting, weather, or occlusions, making it particularly suitable for outdoor applications. Moreover, millimeter-wave radar is cost-effective and easier to deploy on a large scale. Therefore, the fusion of millimeter-wave radar and RGB-D sensors holds significant practical value.

Due to the relatively sparse point clouds generated by millimeter-wave radar and its limited ability to capture scene details compared to optical sensors, existing multi-sensor fusion datasets often overlook the crucial role of millimeter-wave radar in environmental perception. Many current datasets rely on lidar or lower-resolution millimeter-wave radar, which limits their effectiveness. With continuous advancements in millimeter-wave radar technology, higher-performance radar devices have demonstrated greater potential for practical applications. Additionally, many large-scale datasets, constrained by data acquisition methods, do not provide raw millimeter-wave radar data, hindering the full exploitation of radar data in perception tasks. To address this gap, this paper introduces a multi-sensor fusion dataset using a 12Tx-16Rx TI AWR2243 high-resolution millimeter-wave radar sensor for data collection. This dataset provides both raw radar matrices and point cloud data, offering reliable support for in-depth research on millimeter-wave radar.

The goal of this paper is to construct a multi-sensor dataset that includes both RGB-D data and millimeter-wave radar data, providing a comprehensive and rich testing platform for tasks such as depth estimation and environmental perception. Specifically, our dataset encompasses a variety of scenes, including indoor and outdoor environments , ranging from complex laboratory settings, corridors, and office areas to streets and outdoor parking lots. This dataset serves not only as a research platform for depth estimation but also as practical training and testing data for autonomous driving systems and robotic perception systems.

The millimeter-wave radar data in the dataset includes raw radar reflection signals. By processing these signals, we are able to extract high-quality point cloud data. When fused with RGB images and depth maps, this point cloud data provides a more comprehensive perspective for environmental perception tasks such as depth estimation and 3D reconstruction. By combining RGB-D data with millimeter-wave radar point clouds, we can construct a more accurate 3D model, enhancing scene understanding and enabling more precise environmental perception in challenging scenarios and poor lighting conditions.

Building upon the construction of the dataset, we conducted additional experiments by fine-tuning open-source depth estimation models on our dataset to evaluate its performance and reliability in depth estimation tasks. By selecting state-of-the-art depth estimation models, we sought to assess their performance in various environmental conditions and evaluate the dataset's adaptability and robustness in complex, real-world settings. Our contributions are as follows:

1) Constructed a multi-sensor fusion dataset that includes RGB-D and high-quality millimeter-wave radar data, containing both point clouds and raw radar matrices.

2) Proposed and validated a depth completion method based on relative depth estimation and absolute scale supervision.

3) Developed a multi-scenario depth estimation benchmark platform with high practical value.

## II. Related Work

### A. Multi-Sensor Datasets in Autonomous Driving

Multi-sensor fusion technology has been widely applied in fields such as autonomous driving, robot navigation, and indoor positioning, playing a crucial role in a wide range of perception tasks. Existing datasets in this domain are mostly focused on the autonomous driving, covering different sensor configurations and data types. For example, datasets such as Vistas[1], Cityscapes[2], and BDD100K[3] mainly include visual data such as RGB images. In contrast, datasets like KITTI[4], KAIST[5], AS lidar, and ARKitScenes[6] incorporate additional sensors like lidar, IMU, and some thermal imaging and RGB-D data. These datasets are primarily designed for object detection and classification tasks but do not address the acquisition and fusion of millimeter-wave radar data. Another category of datasets, including Astyx[7], nuScenes[8], PixSet[9], RadarScenes[10], TJ4DradSet[11], and NTU4DradLM[12], contains millimeter-wave radar point cloud data; however, these datasets do not provide raw radar data, limiting their support for multi-sensor fusion research based on raw radar data. Additionally, datasets like RADIal[13] and Dual Radar[14], while containing raw millimeter-wave radar data, but do not include visual depth data, restricting their applicability in multi-sensor fusion studies.

TABLE I.   PARAMETERS OF THE RADAR SENSORS

| Parameters | Value |
|---|---|
| Framerate | 5Hz |
| Waveform | FMCW |
| TX antennas | 12 |
| RX antennas | 16 |
| Samples | 128 |
| Angle sampling | 86 |
| Pulse repetition | 64 |
| Max range | 15m |
| Range resolution | 0.117m |
| Max Doppler velocity | 2.02m/s |
| Doppler velocity resolution | 0.254m/s |
| Azimuth resolution | 1.05° |
| Elevation resolution | 22.5° |

### B. Multi-Sensor Datasets in Robot Navigation and Indoor Localization

In the fields of robot navigation and indoor positioning, the Coloradar[15] dataset includes two types of millimeter-wave radar, 3D lidar, and IMU sensors, capturing high-precision real-world data through drill rig poses, supporting multi-sensor fusion-based localization and mapping tasks. However, this dataset does not provide image data. The RadarEyes[16] dataset offers a much larger volume of indoor and outdoor scene data compared to Coloradar, covering various complex environments, but it also lacks visual depth data. The MiliPoint[17] dataset provides millimeter-wave radar and video keypoint data, primarily aimed at human activity recognition research, but similarly lacks visual depth data. The RaDICaL[18] dataset includes raw radar data, RGB-D data, and IMU data, covering various indoor and outdoor scenes, but the millimeter-wave radar used in this dataset has lower spatial resolution.

To address the aforementioned issues, we have constructed the RadarRGBD dataset, which provides over 2,700 frames of diverse indoor and outdoor samples, including millimeter-wave radar ADC data, radar point cloud, and RGB-D images. The millimeter-wave radar we used features up to 86 virtual elements, with an angular resolution close to 1°. During data collection, the system remained stationary, ensuring temporal and spatial synchronization between the radar and image data. These parameter settings and the data collection methodology provide reliable data support for multi-sensor fusion tasks.

## III. The RadarRGBD Dataset

### A. Sensors and Platform

The data collection platform we used is shown in Figure 1. The system consists of a trolley platform and a laptop that connects all the sensors. The laptop is configured with a 12th Gen Intel(R) Core(TM) i7-12700H 2.30 GHz processor, 16GB RAM, 512GB SSD, and running Windows 11. All devices are powered by LiPo batteries. The sensors used include:

- Cascaded millimeter-wave radar sensor: Texas Instruments MMWCAS-RF-EVM (detailed parameters are shown in TABLE I). This radar sensor is used in conjunction with the MMWCAS-DSP-EVM to capture and store the raw radar data.

- Kinect V2 depth camera: Based on Time-of-Flight (ToF) technology, it can simultaneously capture infrared images and depth maps. The optical image resolution is 1920×1080 pixels, with a field of view (FoV) of 84.1°×53.8°; the depth image resolution is 512×424 pixels, with a FoV of 70°×60°, and the depth measurement range is 0.5-8m. The maximum frame rate is 30Hz.

### B. Camera Calibration

For multi-sensor data fusion, sensor calibration is an essential step. Calibration not only improves the measurement accuracy of individual sensors but also aligns the data from different sensors into a common coordinate system, enhancing the accuracy of multi-sensor data fusion.

Since the resolution and field of view of the RGB camera and infrared camera of the Kinect are different, and conventional stereo camera calibration cannot be applied, we use a method where the intrinsic parameters of the RGB camera and infrared camera are calibrated separately first, and then joint calibration is performed to obtain the extrinsic parameters. First, we use the well-known Zhang's[19] calibration method to obtain the intrinsic parameters $\mathbf{A}_{rgb}$ for the RGB camera and $\mathbf{A}_{ir}$ for the infrared camera. During the calibration process, the extrinsic parameters of each camera to the same world coordinate system can be obtained from the images of the calibration board, and then the coordinate transformation can give:

$$P_C = \mathbf{R}_{ir2rgb} * P_{IR} + T_{ir2rgb} \quad (1)$$

where $P_C$ and $P_{IR}$ are the coordinates of a point in the camera coordinate system and the infrared camera coordinate system, respectively, $\mathbf{R}_{ir2rgb}$ and $T_{ir2rgb}$ are the extrinsic parameters from the infrared camera coordinate system to the optical camera coordinate system. To convert the coordinates of the point to the world coordinate system, we have:

$$\mathbf{R}_{rgb} P_W + T_{rgb} = \mathbf{R}_{ir2rgb}\left(\mathbf{R}_{ir} P_W + T_{ir}\right) + T_{ir2rgb} \quad (2)$$

where $\mathbf{R}_{rgb}$ and $T_{rgb}$ are the extrinsic parameters from the world coordinate system to the camera coordinate system, $\mathbf{R}_{ir}$ and $T_{ir}$ are the extrinsic parameters from the world coordinate system to the infrared camera coordinate system, and $P_W$ is the coordinates of a point in the world coordinate system. For each pair of calibration images, we have:

$$\begin{cases} \mathbf{R}_{ir2rgb} = \mathbf{R}_{rgb}\left(\mathbf{R}_{ir}\right)^{-1} \\ T_{ir2rgb} = T_{rgb} - \mathbf{R}_{ir2rgb} T_{ir} \end{cases} \quad (3)$$

In the process of intrinsic parameter calibration, we can obtain the extrinsic parameters for each calibration board. The world coordinate system for each calibration board is confined to a specific corner of the board. For the same pair of calibration boards, their world coordinate system is the same. By calculating the extrinsic parameters between the two sensors for all calibration board images and averaging the results, we can obtain the final extrinsic parameters.

### C. Radar Calibration

Millimeter-wave radar calibration includes antenna channel calibration and extrinsic parameter estimation. The AWR2243 is a 4-chip cascade system, consisting of one master device and three slave devices. Due to inter-chip differences and antenna coupling, the phase channels of these devices are not perfectly aligned. We use a TDM-MIMO form to scan a corner reflector scene at a distance of 5 meters, selecting the first channel as the reference. The frequency compensation for the other channels is calculated as follows:

$$\vec{F} = 2\pi \times \Delta P \times \frac{f_s^{calib}}{f_s^{chirp}} \times \frac{f_{chirp}}{f_{calib}} \bigg/ (N \times I_{interp}) \times \vec{n} \quad (4)$$

where $\Delta P$ is the difference in the Fourier transform peak index of the virtual channel, $f_s^{calib}$ is the signal sampling rate used for compensation, $f_s^{chirp}$ is the signal sampling rate, $f_{chirp}$ is the original frequency modulation slope, $f_{calib}$ is the calibrated frequency slope, $N$ is the number of samples, $\vec{n}$ is the ADC sample index, ranging from 0 to N−1, $I_{interp}$ is the compensation factor, representing the ratio of FFT points to the number of samples.

Phase calibration also uses the first channel as the reference:

$$C_{res} = C_{ref} \big/ C_{other} \quad (5)$$

Where $C_{ref}$ is the complex value of the peak index of the reference virtual channel, $C_{other}$ is the complex value of the peak index of another virtual channel.

For the calibration of the extrinsic parameters of the millimeter-wave radar, multiple retro-reflectors are placed in an open-field scenario. A series of samples are captured using a custom-built system. The pixel coordinates $[u_{ir}, v_{ir}, 1]^T$ and the corresponding depth value $d_{ir}$ can be obtained from the depth map, while the 3D coordinates $P_{Radar}$ of the retro-reflector in the millimeter-wave radar coordinate system are derived from the radar point cloud. Let the extrinsic parameters from the millimeter-wave radar to the infrared camera be represented by $R_{radar2ir}$ and $T_{radar2ir}$, and the coordinates of the retro-reflector in the infrared camera coordinate system be $P_{IR}$, then $[u_{ir}, v_{ir}, 1]^T = \mathbf{A}_{ir} P_{IR} / d_{ir}$. From this, $P_{IR}$ can be obtained. The transformation of the same retro-reflector target between the infrared camera and the millimeter-wave radar coordinate systems is as follows:

$$P_{IR} = \mathbf{R}_{radar2ir} P_{Radar} + T_{radar2ir} \quad (6)$$

The optimization objective is:

$$\arg\min_{R,T} \sum_{i=1}^{N} \left\| R_{radar2ir} P_{Radar} + T_{radar2ir} - P_{IR} \right\|^2 \quad (7)$$

Multiple frames of retro-reflector samples are captured from different angles. The extrinsic parameters between the millimeter-wave radar and the infrared camera coordinate systems can be estimated through least-squares optimization.

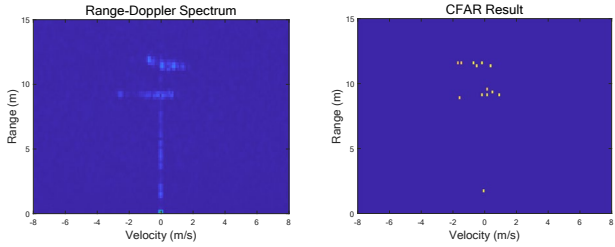

Figure 2. Range-Doppler spectrum(left) and CFAR results(right).

*D. Radar Signal Processing*

Our dataset includes both the ADC radar matrix and point cloud data The radar's transmitting antenna (TX) emits a linear frequency modulated signal (chirp signal), which, when it encounters a target and reflects back, is received by the receiving antenna (RX). The received signal is mixed with the transmitted signal at the receiver, generating an intermediate frequency (IF) signal. The IF signal is then sampled to obtain the radar's raw ADC data cube. The data is collected by the millimeter-wave radar AWR2243, which is equipped with 12 transmitting antennas and 16 receiving antennas, enabling 86 effective virtual channels through MIMO technology.

By applying a one-dimensional fast Fourier transform (1D-FFT) to the "fast-time dimension" data, the distance to the target can be calculated. For multiple continuous chirp signals, a two-dimensional fast Fourier transform (2D-FFT) is performed on the "slow-time dimension" to generate a range-Doppler spectrum, as shown in Figure 2. To detect peaks in the range-Doppler spectrum, a constant false alarm rate (CFAR) detection algorithm is typically used. However, the CFAR process may lose a significant amount of information. To enable other researchers to fully utilize all the information in the radar data, our database provides the raw ADC data. For each detected target point, direction-of-arrival (DOA) estimation is performed on the virtual receive antenna channel dimension. By combining geometric relationships, we can obtain the target's range and azimuth information.

*E. Dataset Structure*

The scenes from which our dataset is collected include indoor laboratory environments, corridors, outdoor parking lots, and street scenes. The data was collected during both sunny daytime and nighttime conditions. The dataset types include raw ADC matrix data from the millimeter-wave radar, radar point cloud data, RGB images, depth images, and infrared images, totaling 2,700 frames of data. We compared RadarRGB with some of the currently most commonly used multi-sensor datasets, and the results are shown in TABLE II:

All the sensor data is stored in a folder named RadarRGBD, which is divided according to data modalities. For ease of use, the millimeter-wave radar ADC data is directly processed into a Radar Cube format, with each frame being a complex array of size (128×86×64), representing the distance, angle, and Doppler dimensions of the sampled points.

The file structure of the collected dataset is shown in Figure 3. Under the RadarRGBD folder, there are six subfolders: five for storing the sensor-collected data, and one for storing the calibration parameters (both intrinsic and extrinsic). The timestamps for each sensor's collected data are stored in a file named timestamps.txt.

TABLE II. DATASET COMPARISON

| Dataset | Radar Type | Radar Formats | Camera | Indoor | Outdoor |
|---|---|---|---|---|---|
| nuScenes | LR | PC | RGB | × | √ |
| PixSet | LR | PC | RGB | × | √ |
| RadarScenes | LR | PC | RGB | × | √ |
| RaDICaL | LR | ADC | RGB-D | √ | √ |
| RADIal | HR | ADC, PC | RGB | × | √ |
| Coloradar | HR | ADC, PC | / | √ | √ |
| TJ4DradSet | HR | PC | RGB | × | √ |
| NTU4DradLM | HR | PC | RGB | × | √ |
| Dual Radar | HR | PC | RGB | × | √ |
| RadarEyes | LR | ADC, PC | Stereo | √ | √ |
| MiliPoint | LR | PC | Stereo | √ | × |
| **Ours** | **HR** | **ADC, PC** | **RGB-D** | **√** | **√** |

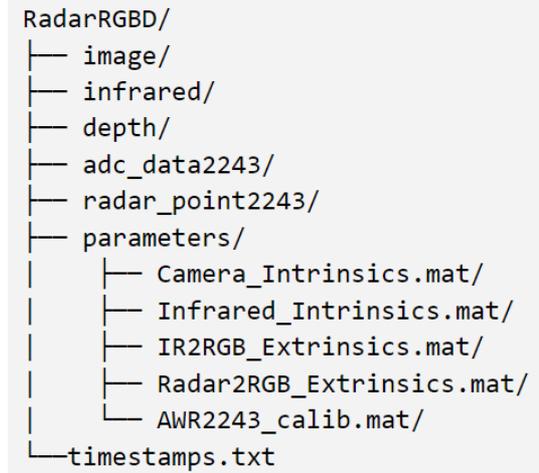

Figure 3. Dataset Structure.

IV. ALGORITHM

In the depth maps obtained from the Kinect V2 device, significant depth gaps are commonly observed. As shown in Figure 4, these gaps are primarily caused by factors such as differences in sensor resolution, field of view (FoV) angles, variations in surface reflectivity, and projection limitations. To fill these gap regions, this paper fine-tunes a pre-trained model based on DepthAnythingV2[20], incorporating absolute depth information from the dataset for global scale supervision. In the depth maps obtained from the Kinect V2 device, significant depth gaps are commonly observed. As shown in Figure 4, these gaps are primarily caused by factors such as differences in sensor resolution, field of view (FoV) angles, variations in surface reflectivity, and projection limitations. To fill these gap regions, this paper fine-tunes a pre-trained model based on DepthAnythingV2[1], incorporating absolute depth information from the dataset for global scale supervision.

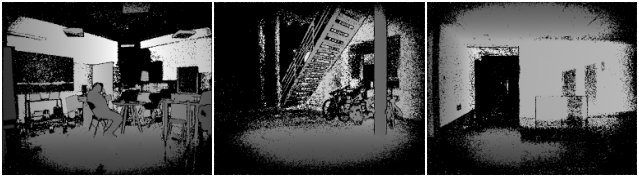

Figure 4. Partial Kinect Captured Depth Map Display.

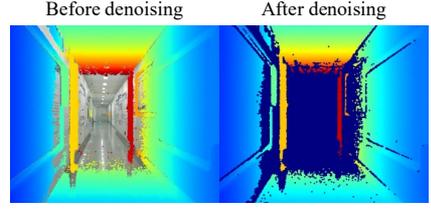

Figure 6. Depth Map Before Denoising (Left) and After Denoising (Right).

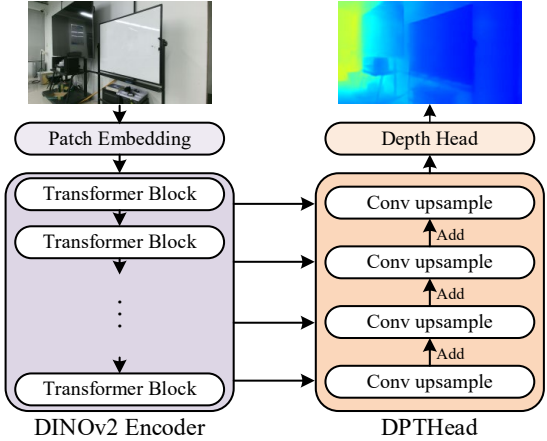

Figure 5. Pre-trained Model Architecture.

The pre-trained model we use is based on the Transformer[1] architecture, utilizing the DINOv2[1] encoder and the DPT[1] depth regression head as the decoder, as shown in Figure 5. Through pre-training on a large-scale synthetic dataset for relative depth estimation using the DINOv2 model with the most parameters, the model can generate pseudo-relative depth maps with good edge detail predictions for real images. Subsequently, DepthAnythingV2 is trained on its generated large-scale image and pseudo-depth pairs to achieve monocular depth estimation with both good detail texture and generalization performance. Our work focuses on metric depth estimation based on its pre-trained model.

Supervising with only sparse ground truth cannot achieve accurate depth estimation in most of the missing regions. Considering this, we use the pre-trained model to generate pseudo-relative depth maps on the RadarRGBD dataset, which are only used to supervise the depth scale of the model and optimize the edges. We employ the SiLog loss function, which optimizes the model based on proportional errors in depth values, helping to mitigate issues arising from scale differences in depth estimation. At valid depth points, we use the smooth L1 loss function for supervising the absolute depth values. To ensure more reasonable depth values are regressed in the missing regions, we use pseudo-relative depth labels generated by the inference of an open-source relative depth estimation model. For the valid depth points, we compute the scaling factor between the model's predictions and the pseudo-relative depth labels. Then, for the remaining points, the labels are scaled and the L1 loss function is applied between the scaled labels and the predicted depth.

$$\mathcal{L}_{ssi}(\hat{d}, d^*) = \mathcal{L}_{l_1-loss}(\hat{d}, z^*)$$
$$z^* = s \cdot d^* + t \quad (8)$$
$$\mathcal{L}_{l_1-loss}(\hat{y}, y) = \frac{1}{N}\sum_{i}^{N}|\hat{y}_i - y_i|$$

Where $s$ and $t$ are the scaling and translation factors, which can be solved using least squares optimization, $\hat{d}$ is the depth predicted by the model, and $d^*$ is the pseudo-relative depth label. To refine the depth estimation at the edges as much as possible, we employ the gradient matching loss function $\mathcal{L}_{GM}$ and the gradient regression loss $\mathcal{L}_{GR}$, as shown below:

$$R = \rho(\hat{z}, z^*)$$
$$\mathcal{L}_{GM} = \frac{1}{N}\sum(\nabla_x R + \nabla_y R) \quad (9)$$
$$\mathcal{L}_{GR} = f_{mse}(E_x(\hat{z}), E_x(z^*)) + f_{mse}(E_y(\hat{z}), E_y(z^*))$$

where $\rho$ is the difference function, which takes the fourth power of the difference between the two predicted values to better correct larger loss discrepancies. $E_x$ and $E_y$ extract the edge information in the two directions, respectively. The final loss function is as follows:

$$\mathcal{L} = \mathcal{L}_{silog} + \mathcal{L}_{ssi} + \mathcal{L}_{smoothl_1} + \mathcal{L}_{GM} + \mathcal{L}_{GR} \quad (10)$$

In some relatively open scenes, the depth maps captured by Kinect are also affected by device and environmental noise, which further impacts the accuracy of depth estimation. Therefore, during the model training process, we introduced morphological operations such as opening and closing from image processing to effectively remove scattered noise points. We applied a closing operation on the depth scatter points to connect large, valid depth scatter areas, preventing them from being incorrectly identified as noise, while ensuring that the variation in noise points remains within an acceptable range. A subsequent opening operation is used to further eliminate noise signals in the sparse regions of the image, resulting in a more precise noise map. This provides a clear supervision signal for training. The removal effect is shown in Figure 6:

The final output depth map is processed through a Sigmoid activation function, mapping it to the range of [0, 1]. After adjusting for depth scale, the output depth values can more accurately reflect the actual measured depth.

V. EXPERIMENTAL RESULTS

Similar to the training fine-tuning strategy of DepthAnythingV2, we ignore the top 10% of the maximum

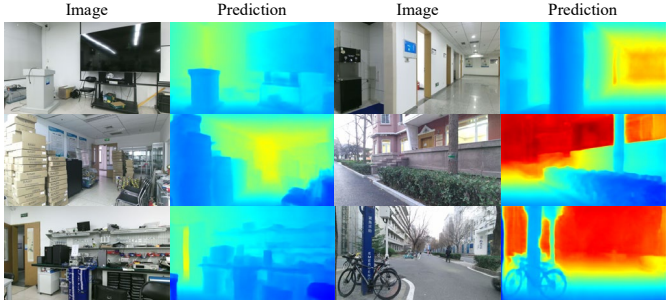

Figure 7. Partial Depth Estimation Results Display.

TABLE III. QUANTITATIVE COMPARISON ON DEPTH COMPLETION METRICS

| Model | RMSE ↓ | MAE ↓ | iRMSE ↑ | iMAE ↑ |
|---|---|---|---|---|
| DepthAnythingV2-vits | 0.136 | 0.086 | 3.622 | 0.080 |
| DepthAnythingV2-vitb | 0.129 | 0.087 | 3.621 | 0.080 |
| DepthAnythingV2-vitl | 0.122 | 0.083 | 3.620 | 0.080 |

loss regions during training, as these areas are likely to exhibit noise labels. All input images are cropped and resized to a consistent size of 518×518 to ensure data consistency. After training for 40 epochs on an NVIDIA V100 GPU, the hole filling results are shown in Figure 7. We trained on ViT models of varying sizes and used some commonly used metrics in depth estimation. The quantitative analysis results for training and inference under the same parameters are presented in TABLE III:

Our modifications to the loss function effectively reduced the depth estimation error to some extent, while leveraging pseudo-relative depth estimation maps allowed us to more reasonably predict the missing ground truth parts in the dataset. Through this multi-stage training process based on deep learning, we aim to effectively fill the holes in depth maps in complex environments and achieve robust depth estimation across various scenes. Our approach not only improves the accuracy of depth completion but also provides an effective solution for data processing and model training in similar tasks in the future.

## VI. CONCLUSIONS

This paper introduces the RadarRGBD dataset, which combines RGB-D data with millimeter-wave radar measurement data, covering various indoor, outdoor, and low-light environments. Notably, this dataset provides high-quality radar point clouds and raw data, offering an essential resource for research on multi-sensor fusion in perception tasks. Additionally, we addressed the hole issue in Kinect V2 depth images by fine-tuning an open-source pre-trained model and improved the depth estimation accuracy through both absolute and relative depth supervision. Experimental results validate the effectiveness of our approach. The RadarRGBD dataset and related methods will provide valuable data and technical support for multi-sensor fusion research, including but not limited to areas such as 3D object detection and scene reconstruction.